\documentstyle[preprint,aps,epsf]{revtex}
\begin{document}
\draft
\title{Conformal invariance in 2-dimensional Discrete Field Theory}
\date{\today }
\author{Serge Winitzki\thanks{%
email: swinitzk@cosmos2.phy.tufts.edu}}
\address{Institute of Cosmology, Department of Physics and Astronomy,\\
Tufts University, Medford, MA 02155}
\maketitle

\begin{abstract}
A discretized massless wave equation in two dimensions, on an appropriately
chosen square lattice, exactly reproduces the solutions of the corresponding
continuous equations. We show that the reason for this exact solution
property is the discrete analog of conformal invariance present in the
model, and find more general field theories on a two-dimensional lattice
that exactly solve their continuous limit equations. These theories describe
in general non-linearly coupled bosonic and fermionic fields and are similar
to the Wess-Zumino-Witten model.
\end{abstract}

\pacs{03.20, 46.10}

\section{Introduction}

Since our spacetime may ultimately prove to be discrete, field theories on
discrete spacetimes have developed an interest in and of themselves \cite
{DiscreteST}. In particular, field theories on two-dimensional discrete
spacetimes can be interpreted as string theories with a discrete string
worldsheet and a continuous target space. Exact solutions for such theories
can often be obtained even after quantization (see, for example, \cite
{DiscreteStrings}). Yet another approach to discretizing spacetime and
constructing a theory of gravity based on stochastic properties of random
lattices is advocated by R. Sorkin \cite{Sorkin}. Investigation of field
theories on discrete spacetime might therefore contribute to our
understanding of discrete quantum picture of the Universe.

We shall focus on the connection between the discrete-spacetime and the
continuous-spacetime versions of the same field theory. This work was
motivated by the (somewhat ``mysterious'') fact noted in Ref.\ \cite{Alex}
that a particular discrete version of the two-dimensional wave equation
gives {\em exact} solutions of the underlying continuous equation. In other
words, the exact solution of the free scalar field theory coincides, on the
lattice points, with the solution of the lattice version of the same theory.
The main intent of this paper is to understand the origin of the exact
equivalence of continuous field theories and their discrete counterparts, as
well as to find more general classes of (classical) field theories that are
exactly solved by properly chosen discretizations. The origin of the exact
solution property of discretized equations turns out to be the discrete
conformal symmetry of the models \cite{Henkel}. The exact solution property
of the massless bosonic model has a direct application in numerical
simulations of strings (which was the original context of Ref.\ \cite{Alex}%
). A quantized version of the discrete conformal symmetry may prove useful
in string theory.

The paper is organized as follows. We first present the exact solution
property of the discretized two-dimensional wave equation. Then, in Sec. \ref
{SecDCI}, an analysis of the discrete equations leads us to the concept of
discrete conformal invariance (DCI), and we give general conditions for a
given discrete field equation to possess DCI. In Sec. \ref{SecGen} we
present a general description of discrete models with DCI, and show that
they all yield exact solutions for the corresponding (first and/or
second-order) differential equations; these solutions are always
algebraically factorized. Further examples of equations with DCI based on
Lie groups and semi-groups are presented in Sec. \ref{Examples}. Conclusions
follow in Sec. \ref{SecConcl}.

\section{Exact solution property of the wave equation}

The two-dimensional wave equation came out of numerical modeling of cosmic
strings \cite{Alex}. The evolution equations for the target space variables $%
X^\mu \left( \tau ,\sigma \right) $ are 
\begin{equation}
X_{\tau \tau }^\mu -X_{\sigma \sigma }^\mu =0,  \label{WaveEqu}
\end{equation}
with gauge conditions 
\begin{equation}
\left( X_\tau ^\mu \pm X_\sigma ^\mu \right) ^2=0.  \label{WEGaugeFix}
\end{equation}
Here, the subscripts $\tau $ and $\sigma $ denote differentiation with
respect to world-sheet coordinates.

The string world-sheet can be discretized to a rectangular lattice with
coordinates $\left( \tau ,\sigma \right) $ and steps $\Delta \tau ,$ $\Delta
\sigma $. A common second-order discretization of (\ref{WaveEqu}) would be 
\begin{equation}
\frac{X^\mu \left( \tau +\Delta \tau ,\sigma \right) -2X^\mu \left( \tau
,\sigma \right) +X^\mu \left( \tau -\Delta \tau ,\sigma \right) }{\Delta
\tau ^2}-\frac{X^\mu \left( \tau ,\sigma +\Delta \sigma \right) -2X^\mu
\left( \tau ,\sigma \right) +X^\mu \left( \tau ,\sigma -\Delta \sigma
\right) }{\Delta \sigma ^2}=0.  \label{CommonDiscr}
\end{equation}

However, a special choice of discretization steps, $\Delta \tau =\Delta
\sigma \equiv \Delta $, was adopted in \cite{Alex}, and it was noted that
the resulting discrete equations are not only simpler, 
\begin{equation}
X^\mu \left( \tau +\Delta ,\sigma \right) +X^\mu \left( \tau -\Delta ,\sigma
\right) -X^\mu \left( \tau ,\sigma +\Delta \right) -X^\mu \left( \tau
,\sigma -\Delta \right) =0,  \label{DWaveEqu}
\end{equation}
\begin{equation}
\left[ X^\mu \left( \tau +\Delta ,\sigma \right) -X^\mu \left( \tau ,\sigma
\pm \Delta \right) \right] ^2=0,  \label{DWEGaugeFix}
\end{equation}
but in addition exactly solve the continuous equation (\ref{WaveEqu}).
Namely, one can easily check that the general solution of (\ref{WaveEqu}), 
\begin{equation}
X^\mu \left( \tau ,\sigma \right) =a_{+}^\mu \left( \tau -\sigma \right)
+a_{-}^\mu \left( \tau +\sigma \right) ,  \label{WESol}
\end{equation}
(where $a_{+}^\mu $ and $a_{-}^\mu $ are vector functions of one argument,
which are determined from the appropriate boundary conditions) exactly
satisfies the discretized equation (\ref{DWaveEqu}). The gauge conditions (%
\ref{DWEGaugeFix}) are then reduced to constraints on $a_{\pm }^\mu $, 
\begin{equation}
\left[ a_{\pm }^\mu \left( x+\Delta \right) -a_{\pm }^\mu \left( x-\Delta
\right) \right] ^2=0.  \label{DWESolConstr}
\end{equation}

To try to understand why this happens, let us examine the discretization (%
\ref{DWaveEqu}) in more detail. Expanding (\ref{DWaveEqu}) in powers of the
discretization step $\Delta $, we notice that the discrete equation (\ref
{DWaveEqu}), on solutions of $\Box X=0$, is satisfied to all orders in $%
\Delta $: 
\begin{eqnarray}
X\left( \tau +\Delta ,\sigma \right) +X\left( \tau -\Delta ,\sigma \right)
-X\left( \tau ,\sigma +\Delta \right) -X\left( \tau ,\sigma -\Delta \right)
&=&  \nonumber \\
\left( \Box X\right) \Delta ^2+\sum_{n=2}^\infty \left( \frac{\partial ^{2n}X%
}{\partial \tau ^{2n}}-\frac{\partial ^{2n}X}{\partial \sigma ^{2n}}\right) 
\frac{2\Delta ^{2n}}{\left( 2n\right) !} &=&0,  \label{DWESeries}
\end{eqnarray}
since 
\[
\frac{\partial ^{2n}X}{\partial \tau ^{2n}}-\frac{\partial ^{2n}X}{\partial
\sigma ^{2n}}=\Box ^nX=0. 
\]
Approximation to all orders means exact solution in the following sense. One
can show that 
\begin{equation}
X^\mu \left( \tau ,\sigma \right) =a_{+}^\mu \left( \tau -\sigma \right)
+a_{-}^\mu \left( \tau +\sigma \right) ,  \label{DWESol}
\end{equation}
where $a_{+}^\mu $ and $a_{-}^\mu $ are now arbitrary vector functions of
one discrete argument, is in fact the general solution of (\ref{DWaveEqu}).
Given a boundary value problem for the continuous equation (\ref{WaveEqu}),
we can restrict the boundary conditions to the discrete boundary points and
find the discrete functions $a_{\pm }^\mu $. Then, from the form of the
solution (\ref{DWESol}) it is clear that $X^\mu \left( \tau ,\sigma \right) $
will coincide, on lattice points, with the solution of the original boundary
value problem. This is what we call the exact solution property of the
discretization (\ref{DWaveEqu}).

Like in the continuous case, the lines $\tau \pm \sigma =const$ are
characteristics of Eq.\ (\ref{DWaveEqu}). It is easily checked that the
gauge conditions (\ref{WEGaugeFix}) are compatible with (\ref{DWaveEqu}), in
the sense that if the discretized gauge conditions (\ref{DWEGaugeFix}) hold
at some point $\left( \tau ,\sigma \right) $, then the evolution equation (%
\ref{DWaveEqu}) guarantees that they hold everywhere along the
characteristic that intersects the point $\left( \tau ,\sigma \right) $.
However, the exact solutions of the equations (\ref{WaveEqu})$-$(\ref
{WEGaugeFix}) do not necessarily satisfy (\ref{DWEGaugeFix}) for any
discretization step $\Delta $, because the continuous constraint (\ref
{WEGaugeFix}) imposed on the derivative of $a_{\pm }^\mu $ is non-linear and
does not integrate to a discrete constraint (\ref{DWESolConstr}). In this
sense, the discrete constraints (\ref{DWEGaugeFix}) provide only an
approximate solution of (\ref{WEGaugeFix}).

We have shown that a particular discretization (\ref{DWaveEqu}) of the wave
equation exactly solves\ the corresponding continuum equation, regardless of
the magnitude of the lattice step $\Delta $. The remaining part of this work
is intended to explain and explore this unexpected phenomenon. The
``mysterious'' cancellation of all higher-order terms in (\ref{DWESeries})
is due to conformal invariance, as will be explored in the next two sections.

\section{Discrete conformal invariance of the wave equation}

\label{SecDCI}

Since the exact solution property is independent of the discretization step $%
\Delta $, we are motivated to explore some kind of scale invariance in the
model. We first notice that Eq.\ (\ref{DWaveEqu}) does not mix the values of
the field on ``even'' and ``odd'' sublattices, each sublattice consisting of
points with $\tau +\sigma $ even or odd. Therefore we shall further consider
only one of these two sublattices, which is itself a square lattice in the
lightcone coordinates $\xi _{\pm }\equiv \tau \pm \sigma $. In Fig.\ 1 we
drew four adjacent squares, each square corresponding to a group of four
adjacent points on the lightcone lattice with the field values $X^\mu $
related by (\ref{DWaveEqu}), for instance 
\[
X^\mu \left( a^{\prime }\right) +X^\mu \left( b^{\prime }\right) -X^\mu
\left( c^{\prime }\right) -X^\mu \left( c\right) =0. 
\]
By adding together the corresponding equations of motion for all four
squares in Fig.\ 1, it is straightforward to show that the same equation
actually holds for points $a$, $b$, $c$, and $d$ that lie at the vertices of
a larger $2\times 2$ square, i.e. 
\begin{equation}
X^\mu \left( a\right) +X^\mu \left( b\right) -X^\mu \left( c\right) -X^\mu
\left( d\right) =0.
\end{equation}

In a similar fashion we find that the discrete equation (\ref{DWaveEqu})
remains valid if we replace the discretization step $\Delta $ by its
multiple $n\Delta $, which means removing all but $n$-th points from the
lattice, and effectively stretching the lattice scale by a factor of $n$. We
could call this fact {\em scale invariance} of the discrete equations. In
fact, an even more general kind of invariance holds: namely, Eq.\ (\ref
{DWaveEqu}) retains its form when applied to the four vertices of any
rectangle on the lightcone lattice (i.e. with sides parallel to the null
directions; such a rectangle would have its sides parallel to the $\xi _{\pm
}$ axes). One can prove this by noticing that, by adding Eq.\ (\ref{DWaveEqu}%
) applied to points $a-e-a^{\prime }-c^{\prime }$ and $a^{\prime }-c^{\prime
}-b^{\prime }-c$, one obtains 
\[
X^\mu \left( a\right) +X^\mu \left( b^{\prime }\right) -X^\mu \left(
e\right) -X^\mu \left( c\right) =0, 
\]
which is the same equation applied to points $a-b^{\prime }-e^{\prime }-c$
which lie at vertices of a rectangle rather than a square. The same
procedure shows that Eq.\ (\ref{DWaveEqu}) applies to points $f-a^{\prime
}-b-c$ as well. After deriving (\ref{DWaveEqu}) for the ``elementary'' $%
2\times 1$ rectangles, it is clear that, for any given rectangle, one only
needs to add together the relation (\ref{DWaveEqu}) for all squares inside
it to obtain the same relation for the vertices of the rectangle.

The fact that Eq.\ (\ref{DWaveEqu}) applies to vertices of any lattice
rectangle means that if we remove all points lying on any number of lines $%
\tau +\sigma =const$ and $\tau -\sigma =const$ from the lattice, the
discrete equations will still apply to the remaining points. In lightcone
coordinates, such a transformation amounts to replacing the (discrete)
coordinates $\xi _{\pm }$ by functions of themselves: 
\begin{equation}
\tilde \xi ^{+}=f^{+}\left( \xi ^{+}\right) ,\quad \tilde \xi
^{-}=f^{-}\left( \xi ^{-}\right) .  \label{DCTrans}
\end{equation}
Here, the functions $f^{\pm }\left( \xi ^{\pm }\right) $ are arbitrary
monotonically increasing discrete-valued functions of one discrete argument.
The monotonicity of these functions is necessary to preserve causal
relations on the spacetime.

We immediately notice a similarity between (\ref{DCTrans}) and conformal
transformations in a two-dimensional pseudo-Euclidean space. As is well
known, the wave equation (\ref{WaveEqu}) allows arbitrary conformal
transformations of the world-sheet coordinates, 
\begin{equation}
\tilde \xi ^{+}=f^{+}\left( \xi ^{+}\right) ,\quad \tilde \xi
^{-}=f^{-}\left( \xi ^{-}\right) .  \label{twodCoordChange}
\end{equation}
This is the most general form of the coordinate transformation that
preserves the pseudo-Euclidean metric 
\begin{equation}
ds^2=d\tau ^2-d\sigma ^2=d\xi _{+}d\xi _{-}
\end{equation}
up to a conformal factor. Transformations (\ref{DCTrans}) are obviously the
discrete analog of conformal transformations (\ref{twodCoordChange}) on the
lightcone lattice.

We have, therefore, found that the discrete wave equation (\ref{DWaveEqu})
is invariant under the discrete conformal transformations (\ref{DCTrans}).
We shall refer to this as the {\em discrete conformal invariance} (DCI) of
the discrete wave equation. The natural question is then whether there exist
other equations with the property of DCI, and whether such equations also
deliver exact solutions of their continuous limits. In the remaining
sections, we will answer this question in the positive, establishing a
relation between the exact solution property and DCI.

Note that the constraint equations (\ref{DWEGaugeFix}) do not preserve their
form under discrete conformal transformations. For instance, if the
constraints hold for some step size $\Delta $, 
\[
\left[ X^\mu \left( \tau +\Delta ,\sigma \right) -X^\mu \left( \tau ,\sigma
+\Delta \right) \right] ^2=0,\quad \left[ X^\mu \left( \tau ,\sigma +\Delta
\right) -X^\mu \left( \tau -\Delta ,\sigma +2\Delta \right) \right] ^2=0, 
\]
it does not follow in general that they also hold for step size $2\Delta $: 
\[
\left[ X^\mu \left( \tau +\Delta ,\sigma \right) -X^\mu \left( \tau -\Delta
,\sigma +2\Delta \right) \right] ^2\neq 0. 
\]
Accordingly, as we have seen in the previous Section, the discrete
constraint equations do not exactly solve their continuous counterparts.

\section{A general form of a DCI field theory}

\label{SecGen}To try to generalize Eq.\ (\ref{DWaveEqu}), we write the
generic discrete evolution equation on a square lightcone lattice as 
\begin{equation}
X\left( \tau +\Delta ,\sigma \right) =F\left[ X\left( \tau ,\sigma +\Delta
\right) ,X\left( \tau ,\sigma -\Delta \right) ,X\left( \tau -\Delta ,\sigma
\right) \right] ,  \label{DEvol}
\end{equation}
where $F$ is an unknown function. As we have found in the previous Section,
the property of DCI will be satisfied if (\ref{DEvol}) is invariant under
the ``elementary'' conformal transformations, that is, if the relation (\ref
{DEvol}) is valid when applied to the vertices of all $2\times 1$ lattice
rectangles. This requirement can be written as two functional conditions on $%
F$: 
\begin{mathletters}
\label{DCIConds}
\begin{eqnarray}
F\left( a,F\left( b,c,d\right) ,b\right) &=&F\left( a,c,d\right) , \\
F\left( F\left( a,c,d\right) ,b,c\right) &=&F\left( a,b,d\right) .
\end{eqnarray}
Here, $a$, $b$, $c$, and $d$ are arbitrary field values which may be scalar
or vector (or even belong to a non-linear manifold of a Lie group, as in our
examples below), and $F$ is a similarly valued function. Before we try to
solve these conditions for $F$, we would like to show that for any function $%
F$ satisfying (\ref{DCIConds}), the discrete evolution equation (\ref{DEvol}%
) exactly solves its continuous limit.

The continuous limit of (\ref{DEvol}) is obtained by expanding it in powers
of $\Delta $, for instance 
\end{mathletters}
\begin{equation}
X\left( \tau ,\sigma +\Delta \right) =X_0+X_\sigma \Delta +\frac{X_{\sigma
\sigma }}2\Delta ^2+O\left( \Delta ^3\right) ,
\end{equation}
and with the assumption that $F\left( X_0,X_0,X_0\right) =X_0$, which is
natural if we suppose that a constant function $X=X_0$ must be a solution of
(\ref{DEvol}), we obtain the following equations corresponding to first and
second powers of $\Delta $: 
\begin{equation}
X_\tau =F_1X_\sigma -F_2X_\sigma -F_3X_\tau ,  \label{ContEqu1}
\end{equation}
\begin{equation}
X_{\tau \tau }=\left( F_1+F_2\right) X_{\sigma \sigma }+F_3X_{\tau \tau
}+\left( F_{11}+F_{22}-2F_{12}\right) X_\sigma X_\sigma +\left(
F_{23}-F_{31}\right) \left( X_\sigma X_\tau +X_\tau X_\sigma \right)
+F_{33}X_\tau X_\tau .  \label{ContEqu2}
\end{equation}
Here, $F_i$ and $F_{ij}$ are derivatives of $F$ with respect to its three
numbered arguments (which are in general vector arguments, but we suppressed
the indices in the above equations). One can easily verify that for $F\left(
a,b,c\right) =a+b-c$, these equations give the usual wave equation (\ref
{WaveEqu}).

The derivatives $F_i$ and $F_{ii}$ are constrained by Eqs.\ (\ref{DCIConds}%
). For example, the first derivatives satisfy 
\begin{equation}
F_1=F_1F_1,\quad F_2=F_2F_2,\quad F_3=-F_1F_2,  \label{FirstDerivs}
\end{equation}
where $F_i$ are understood as linear operators on the tangent target space.
It follows that $F_i$ are projection operators (i.e. operators $P$ that obey 
$P^2=P$) with eigenvalues $0$ and $1$ only.

Now we shall show that the equations (\ref{ContEqu1})$-$(\ref{ContEqu2}) are
exactly solved by (\ref{DEvol}) if the DCI\ conditions (\ref{DCIConds})
hold. Denote the exact solution of the continuous equations (\ref{ContEqu1})$%
-$(\ref{ContEqu2}) by $X_e\left( \tau ,\sigma \right) $, and the solution of
the discrete equation (\ref{DEvol}) by $X_d\left( \tau ,\sigma \right) $.
Here, $\tau $ and $\sigma $ are discrete lattice coordinates, and we assume
that $X_e=X_d$ on the lines $\tau \pm \sigma =0$ (Fig.\ 2). Since Eqs.\ (\ref
{ContEqu1})$-$(\ref{ContEqu2}) were obtained from (\ref{DEvol}) by expansion
in $\Delta $ up to third-order terms, $X_e$ satisfies the discrete equation (%
\ref{DEvol}) up to $O\left( \Delta ^3\right) $, i.e. 
\begin{equation}
\left. X_d-X_e\right| _{\tau =2\Delta ,\sigma =0}=C\left( \Delta \right)
\Delta ^3,\quad C\left( \Delta \right) =C_0+O\left( \Delta \right) ,
\label{CubeError}
\end{equation}
in the limit of small $\Delta $. Now we shall scale up the lattice by an
arbitrarily chosen factor of $n$. As we have seen in the previous Section,
from DCI it follows that Eq.\ (\ref{DEvol}) applies also to the vertices of
an $n\times n$ square on the lightcone lattice, and this is a consequence of
using Eq.\ (\ref{DEvol}) $n^2$ times, once for each ``elementary'' square.
At each elementary square, we can replace the field values $X_d$ by $X_e$
and introduce an error of $C\left( \Delta \right) \Delta ^3$. An error of $%
\delta X$ in $X$ entails an error of not more than $3\delta X$ in $F\left(
X,X,X\right) $, because the first derivatives of $F$ are operators with
eigenvalues of $0$ and $1$ only. Therefore, replacing $X_d$ by $X_e$ in the $%
n\times n$ square entails an error in $X\left( \tau =2n,\sigma =0\right) $
of not more than $3n^2C\left( \Delta \right) \Delta ^3$: 
\begin{equation}
\left. X_d-X_e\right| _{\tau =2n\Delta ,\sigma =0}\leq 3n^2C\left( \Delta
\right) \Delta ^3.
\end{equation}
However, we can also apply (\ref{CubeError}) directly to the $n\times n$
square to obtain 
\begin{equation}
\left. X_d-X_e\right| _{\tau =2n\Delta ,\sigma =0}=C\left( n\Delta \right)
n^3\Delta ^3.
\end{equation}
It means that 
\[
C\left( n\Delta \right) \leq \frac 3nC\left( \Delta \right) , 
\]
which forces $C\left( \Delta \right) =0$ since $C\left( \Delta \right) $ is
a polynomial in $\Delta $. Therefore, the approximation (\ref{CubeError}) is
in fact exact.

This argument shows that it was in fact only necessary to expand (\ref{DEvol}%
) up to second order in $\Delta $, and all higher-order terms will lead to
differential equations which are consequences of (\ref{ContEqu1})$-$(\ref
{ContEqu2}), just as we have seen in the case of the wave equation. It also
shows that the differential equations corresponding to given DCI field
equations (\ref{DEvol}) are always second-order or lower.

Using the evolution equation (\ref{DEvol}), one can write the exact solution
of the continuous equations for boundary conditions on the lightcone $\tau
\pm \sigma =0$. To find $X\left( \tau ,\sigma \right) $ at a point $\left(
\tau ,\sigma \right) $ within the future lightcone of the origin, we
construct a lattice that has $\left( \tau ,\sigma \right) $ as one of its
points, and then apply Eq.\ (\ref{DEvol}) to the rectangle $\left( \tau
,\sigma \right) -\left( \frac{\tau +\sigma }2,\frac{\tau +\sigma }2\right)
-\left( \frac{\tau -\sigma }2,-\frac{\tau -\sigma }2\right) -\left(
0,0\right) $ and explicitly write $X\left( \tau ,\sigma \right) $ through
the boundary values: 
\begin{equation}
X\left( \tau ,\sigma \right) =F\left( X\left( \frac{\tau +\sigma }2,\frac{%
\tau +\sigma }2\right) ,X\left( \left( \frac{\tau -\sigma }2,-\frac{\tau
-\sigma }2\right) \right) ,X\left( 0,0\right) \right) .  \label{ExactSol}
\end{equation}

Now we try to describe a general class of functions $F$ satisfying (\ref
{DCIConds}). First, we find by combining Eqs.\ (\ref{DCIConds}) that 
\begin{equation}
F\left( a,F\left( d,b,c\right) ,c\right) =F\left( F\left( a,d,c\right)
,b,c\right) .
\end{equation}
If we denote 
\begin{equation}
a*_cb\equiv F\left( a,b,c\right) ,  \label{Multi}
\end{equation}
the above will look like an associative law for a binary operation $*_c$: 
\begin{equation}
a*_c\left( d*_cb\right) =\left( a*_cd\right) *_cb.
\end{equation}
The condition $F\left( a,a,a\right) =a$ looks like a unity law 
\begin{equation}
a*_aa=a,
\end{equation}
although it doesn't follow from (\ref{DCIConds}) that $a*_ab=b$ for all $b$.
However, if we suppose that (\ref{Multi}) is actually an operation of group
multiplication, with the usual unity law 
\begin{equation}
a*_ab=b\text{ for all }b,
\end{equation}
and the inverse operation $b^{-1}$, then it is shown in the Appendix that
there exists a reparametrization $r$ of the field values such that the
operation $*_c$ is written as 
\begin{equation}
r\left( a*_cb\right) =r\left( a\right) *r\left( c^{-1}\right) *r\left(
b\right) ,  \label{GenM}
\end{equation}
where by $*$ we denote the group multiplication in the group of field
values. Note that although Eq.\ (\ref{GenM}) is not symmetric with respect
to interchange of $a$ and $b$, such interchange is equivalent to a
reparametrization $r\left( a\right) =a^{-1}$ (where $a^{-1}$ is the group
inverse of $a$).

We arrive at a picture of a field equation of the type (\ref{DEvol}),
derived from a group multiplication in an arbitrary Lie group. We shall
assume that any needed reparametrization is already effected, and that the
evolution is directly given by (\ref{DEvol}) with 
\begin{equation}
F\left( a,b,c\right) =a*c^{-1}*b.  \label{GenF}
\end{equation}
For example, if we consider a vector space as an Abelian group with addition
of vectors as the operation $*$, we again obtain the formula 
\begin{equation}
F\left( a,b,c\right) =a-c+b
\end{equation}
for the discrete wave equation.

In view of Eq.\ (\ref{GenF}), the exact solution (\ref{ExactSol}) has an
algebraically factorized form, as a product of functions of the lightcone
coordinates: 
\begin{equation}
X\left( \tau ,\sigma \right) =a\left( \tau +\sigma \right) *b\left( \tau
-\sigma \right)  \label{ESolGroup}
\end{equation}
(the constant $c^{-1}$ is absorbed by either of the functions). This is a
characteristic feature of the DCI equations we are concerned with.

The continuous limit of the discrete equations (\ref{DEvol})$-$(\ref{GenF})\
can also be described in terms of an arbitrary Lie group $G$ as a target
space; it is a Wess-Zumino-Witten (WZW)-type model \cite{WZW}. If $X$ is a
field with values in a (matrix) group $G$ defined on a two-dimensional
manifold $M$, the action functional of the WZW model is defined by 
\begin{equation}
L=\frac 1{2\lambda }\int_MTr\left( X^{-1}\partial _aX\cdot X^{-1}\partial
^aX\right) d^2M+\frac k{24\pi }\int_B\epsilon ^{abc}Tr\left( X^{-1}\partial
_aX\cdot X^{-1}\partial _bX\cdot X^{-1}\partial _cX\right) d^3B,
\end{equation}
where $B$ is an auxiliary $3$-dimensional manifold whose boundary is $M$,
and $Tr$ is the matrix trace operation. For a specific choice $k=\pm \frac 4%
\lambda $, the equations of motion in the lightcone coordinates become 
\begin{equation}
\partial _{\mp }\left( X^{-1}\partial _{\pm }X\right) =0,
\end{equation}
with the general solution 
\begin{equation}
X\left( \xi _{+},\xi _{-}\right) =X_{+}\left( \xi _{+}\right) \cdot
X_{-}\left( \xi _{-}\right) \text{ or }X_{-}\left( \xi _{-}\right) \cdot
X_{+}\left( \xi _{+}\right) ,  \label{WZWSol}
\end{equation}
with $X_{\pm }$ being arbitrary $G$-valued functions. It is immediately seen
that the solution (\ref{WZWSol}) is the same as (\ref{ESolGroup}), if we
(naturally) choose $*$ to be the group multiplication in $G$.

\section{Examples of field theories with DCI}

\label{Examples}In this Section we present some examples to illustrate the
constructions of Sec. \ref{SecGen} and to give a physical interpretation of
the field theories arising from them.

\subsection{Interacting scalar fields}

As was noted in the previous section, we can obtain the discrete wave
equation by using the general formula (\ref{GenF}) on a vector space
considered as an additive group of vectors. Since all commutative Lie groups
are locally isomorphic to a vector space, we should take a non-commutative
group to find a less trivial example. The simplest non-commutative Lie group
has two parameters and can be realized by matrices of the form 
\begin{equation}
\left\{ \alpha ,a\right\} \equiv \left( 
\begin{array}{cc}
e^\alpha & a \\ 
0 & 1
\end{array}
\right) .  \label{MatrixTwod}
\end{equation}
The composition law of the group is 
\begin{equation}
\left( 
\begin{array}{cc}
e^\alpha & a \\ 
0 & 1
\end{array}
\right) \left( 
\begin{array}{cc}
e^\beta & b \\ 
0 & 1
\end{array}
\right) =\left( 
\begin{array}{cc}
e^{\alpha +\beta } & a+e^\alpha b \\ 
0 & 1
\end{array}
\right) ,
\end{equation}
or, written more compactly, 
\begin{equation}
\left\{ \alpha ,a\right\} \left\{ \beta ,b\right\} =\left\{ \alpha +\beta
,a+e^\alpha b\right\} .  \label{twodMC}
\end{equation}
The inverse element is given by $\left\{ \alpha ,a\right\} ^{-1}=\left\{
-\alpha ,-e^{-\alpha }a\right\} $.

A calculation shows that the continuous limit of the discrete evolution
equation (\ref{GenF}) with the composition law (\ref{twodMC}) is (in
lightcone coordinates $\xi _{\pm }$) 
\begin{mathletters}
\label{twodME}
\begin{eqnarray}
\frac{\partial ^2\alpha }{\partial \xi _{+}\partial \xi _{-}} &=&0,
\label{twodMEalpha} \\
\frac{\partial ^2a}{\partial \xi _{+}\partial \xi _{-}} &=&\frac{\partial
\alpha }{\partial \xi _{+}}\frac{\partial a}{\partial \xi _{-}}.
\end{eqnarray}
Note that the non-commutativity of the group leads to asymmetry with respect
to the spatial reflection (i.e. interchange of $\xi _{+}$ and $\xi _{-}$),
but at the same time spatial reflection is equivalent to reparametrization $%
\alpha \rightarrow -\alpha ,$ $a\rightarrow -a\exp \left( -\alpha \right) $
which corresponds to taking the inverse matrix to (\ref{MatrixTwod}).

The exact solution of (\ref{twodME}) is obtained from the general formula (%
\ref{ExactSol}): 
\end{mathletters}
\begin{eqnarray}
\alpha \left( \tau ,\sigma \right) &=&\alpha _{+}\left( \tau +\sigma \right)
+\alpha _{-}\left( \tau -\sigma \right) , \\
a\left( \tau ,\sigma \right) &=&a_{+}\left( \tau +\sigma \right) +e^{\alpha
_{+}\left( \tau +\sigma \right) }a_{-}\left( \tau -\sigma \right) ,
\end{eqnarray}
with arbitrary functions $\alpha _{\pm }$ and $a_{\pm }$.

The equations (\ref{twodME}) were obtained from a two-dimensional group and
describe a pair of coupled scalar fields. More generally, one may start with
an arbitrary $n$-dimensional non-commutative group $G$ and construct the
corresponding discrete and continuous equations describing $n$ coupled
scalar fields. The group structure will then be reflected in the coupling of
the fields: for example, if the group $G$ has a commutative subgroup $H$,
then the corresponding parameters will satisfy a free wave equation. This
can be easily seen from the example above: the elements of the form $\left\{
\alpha ,a=0\right\} $ form a commutative subgroup, and the corresponding
equation (\ref{twodMEalpha}) for $\alpha $ is a wave equation.

\subsection{Fermionic fields: the discrete Dirac equation}

So far, we have been dealing with scalar fields. Now we shall consider the
possibility of DCI equations describing fermions. The 2-dimensional massless
Dirac equation for the two-component fermionic field $\psi \left( \xi
^{+},\xi ^{-}\right) $ is 
\begin{equation}
\gamma ^a\partial _a\psi =0,  \label{twodDirac}
\end{equation}
where the corresponding Dirac matrices satisfy the usual relations of a
Clifford algebra $\left\{ \gamma ^a,\gamma ^b\right\} =2g^{ab}$ and can be
chosen as 
\begin{equation}
\gamma ^{-}=\left( 
\begin{array}{cc}
0 & 0 \\ 
1 & 0
\end{array}
\right) ,\quad \gamma ^{+}=\left( 
\begin{array}{cc}
0 & 1 \\ 
0 & 0
\end{array}
\right) .  \label{twodDiracGammas}
\end{equation}
Here, the metric $g^{ab}$ in the lightcone coordinates $\xi ^a$ is $%
g^{+-}=g^{-+}=1$, $g^{++}=g^{--}=0$.

With the choice (\ref{twodDiracGammas}), Eq.\ (\ref{twodDirac}) becomes 
\begin{equation}
\partial _{+}\psi _1=0,\quad \partial _{-}\psi _2=0,  \label{twodDiracLC}
\end{equation}
where $\psi _{1,2}$ are left- and right-moving components of the field $\psi
\left( \xi ^{+},\xi ^{-}\right) $.

The corresponding lattice equations are 
\begin{mathletters}
\label{twodDiracD}
\begin{eqnarray}
\psi _1\left( \tau +\Delta ,\sigma \right) -\psi _1\left( \tau ,\sigma
+\Delta \right) &=&0, \\
\psi _2\left( \tau +\Delta ,\sigma \right) -\psi _2\left( \tau ,\sigma
-\Delta \right) &=&0.
\end{eqnarray}
Their solution is 
\end{mathletters}
\begin{equation}
\psi _1=f_1\left( \tau +\sigma \right) ,\quad \psi _2=f_2\left( \tau -\sigma
\right) ,
\end{equation}
which is also an exact solution of the continuous equations (\ref
{twodDiracLC}). (Here, $f_{1,2}$ are functions of discrete argument
determined by boundary conditions.)

Since we again discovered a case of an exact solution, we naturally try to
write the lattice equations (\ref{twodDiracD}) in the form (\ref{DEvol})$-$(%
\ref{GenF}). This is possible if we define the multiplication operation $*$
by 
\begin{equation}
\left( 
\begin{array}{c}
\psi _1 \\ 
\psi _2
\end{array}
\right) *\left( 
\begin{array}{c}
\phi _1 \\ 
\phi _2
\end{array}
\right) =\left( 
\begin{array}{c}
\phi _1 \\ 
\psi _2
\end{array}
\right) .
\end{equation}
(Note how the left- and right-handed components of the field propagate to
the left and to the right of the multiplication sign.) Such a multiplication
operation is associative, but does not allow a unity element and is not
invertible, which would make it impossible to write $a*c^{-1}*b$ as in (\ref
{GenF}). However, this operation has the property that $a*x*b=a*b$
regardless of the value of $x$, and therefore we can disregard $c^{-1}$ in (%
\ref{GenF}). A matrix representation of this multiplication operation can be
defined by 
\begin{equation}
\left( 
\begin{array}{c}
\psi _1 \\ 
\psi _2
\end{array}
\right) \equiv \left( 
\begin{array}{cccc}
1 & \psi _1 &  &  \\ 
0 & 0 &  &  \\ 
&  & 1 & 0 \\ 
&  & \psi _2 & 0
\end{array}
\right) .  \label{MatrixSemigroup}
\end{equation}

Since the derivations of Sec. \ref{SecGen} only use the associativity of the
multiplication operation $*$, all our considerations apply also to cases
where this operation does not have an inverse, such as in the case of {\em %
semigroups} \cite{Algebra}. We shall be interested in a specific example of
the semigroup structure represented in (\ref{MatrixSemigroup}), where $\psi
_1$ and $\psi _2$ can be, in general, multi-component fields; we shall refer
to such a structure as a ``fermionic semigroup''.

\subsection{Coupled bosons and fermions}

Heuristically, a Lie group generates bosons and a fermionic semigroup
generates fermions in DCI field theories. The direct product of a group and
a semigroup would result in a theory describing uncoupled bosons and
fermions. An example of a model containing coupled bosons and fermions can
be obtained from a semigroup built as a {\em semi-direct product} of a group
and a fermionic semigroup. A semi-direct (or ``twisted'') product of two
semi-groups $S$ and $G$ can be defined if $S$ {\em acts} on $G$, i.e. if for
each $s\in S$ there is a map $s:G\rightarrow G$ such that 
\begin{equation}
s\left( g_1*g_2\right) =s\left( g_1\right) *s\left( g_2\right)
\label{MapHom}
\end{equation}
and 
\begin{equation}
s_1\left( s_2\left( g\right) \right) =\left( s_1*s_2\right) \left( g\right) .
\label{HomMap}
\end{equation}
(Here, the multiplication denotes the respective semi-group operation in $S$
or $G$, where appropriate.) The semi-direct product of $S$ and $G$ is the
set of pairs $\left\{ s,g\right\} $ with the multiplication defined by 
\begin{equation}
\left\{ s_1,g_1\right\} *\left\{ s_2,g_2\right\} \equiv \left\{
s_1*s_2,g_1*s_1\left( g_2\right) \right\} .  \label{SemiMult}
\end{equation}
One can easily check that this operation is associative. Of course, a group
is also a semigroup, and the semi-direct product construction can be applied
to two groups or to a group and a semigroup as well.

The existence of an associative binary operation on the target space is
really all we need to build a DCI field theory. We can obtain a generic
theory of this kind containing both bosons and fermions by taking the
fermionic semigroup $S$ and some Lie group $G$ on which $S$ acts. Such pairs 
$\left( S,G\right) $ can be constructed for arbitrary dimensions of $S$ and $%
G$. We shall, to illustrate this construction, couple the two previous
examples and arrive to a model with two interacting bosons and three
fermions.

To do this, we take $S$ to be the simplest fermionic semigroup of (\ref
{MatrixSemigroup}). As $G$ we choose a group like one represented by (\ref
{MatrixTwod}), but with two parameters $a_{1,2}$: 
\begin{equation}
\left\{ \alpha ,a_1,a_2\right\} \equiv \left( 
\begin{array}{ccc}
e^\alpha & a_1 & a_2 \\ 
0 & 1 & 0 \\ 
0 & 0 & 1
\end{array}
\right) .
\end{equation}
To couple bosons with fermions, we need to introduce some action of $S$ on $%
G $. For example, we can multiply the column of parameters $a_{1,2}$ of the
group $G$ by the right-moving part of the matrix $s$, 
\begin{equation}
\left( 
\begin{array}{cc}
1 & 0 \\ 
\psi _2 & 0
\end{array}
\right) \left( 
\begin{array}{c}
a_1 \\ 
a_2
\end{array}
\right) =\left( 
\begin{array}{c}
a_1 \\ 
\psi _2a_1
\end{array}
\right) ,
\end{equation}
which defines the action of $\left\{ \psi _1,\psi _2\right\} $ on $\left\{
\alpha ,a_1,a_2\right\} $ as 
\begin{equation}
\left( 
\begin{array}{c}
\psi _1 \\ 
\psi _2
\end{array}
\right) \left( \left\{ \alpha ,a_1,a_2\right\} \right) \equiv \left\{ \alpha
,a_1,\psi _2\alpha _1\right\} .
\end{equation}
One can check that the conditions (\ref{MapHom})$-$(\ref{HomMap}) hold for
such an action. The multiplication law of the resulting five-parametric
semigroup is 
\begin{equation}
\left\{ \psi _1,\psi _2;\alpha ,a_1,a_2\right\} *\left\{ \phi _1,\phi
_2;\beta ,b_1,b_2\right\} =\left\{ \phi _1,\psi _2;\alpha +\beta
,a_1+e^\alpha b_1,a_2+e^\alpha \psi _2b_1\right\}
\end{equation}
with a matrix realization 
\begin{equation}
\left\{ \psi _1,\psi _2;\alpha ,a_1,a_2\right\} \equiv \left( 
\begin{array}{cccccc}
e^\alpha &  &  &  &  &  \\ 
& e^\alpha & 0 & a_1 &  &  \\ 
& e^\alpha \psi _2 & 0 & a_2 &  &  \\ 
& 0 & 0 & 1 &  &  \\ 
&  &  &  & 1 & \psi _1 \\ 
&  &  &  & 0 & 0
\end{array}
\right) .
\end{equation}
The multiplication in this semigroup does not allow an inverse operation.
Nevertheless, the problem with $c^{-1}$ in the general formula (\ref{GenF})
is circumvented because we have chosen the twisting of the semigroup and the
group in such a way as to make the product $a*c*b$ independent of the $\psi
_{1,2}$ components of $c$.

The continuous limit equations in this model are 
\begin{eqnarray*}
\partial _{+-}\alpha &=&0, \\
\partial _{+}\psi _1 &=&0,\quad \partial _{-}\psi _2=0, \\
\partial _{+-}a_1 &=&\partial _{+}a_1\partial _{-}\alpha , \\
\partial _{+}a_2 &=&\psi _2\partial _{+}a_1.
\end{eqnarray*}

As can be seen from the above, $\psi _{1,2}$ and $a_2$ are fermions, while $%
\alpha $ and $a_1$ are bosons, $a_1$ is coupled to $\alpha $ and $a_2$ is
coupled to $\psi _2$ and $a_1$. (Here, we formally refer to the fields with
first-order equations of motion as ``fermionic''.) Similar models can be
constructed for a larger number of coupled bosonic and fermionic fields.
Note that the field $a_2$ which was a boson in the $\{\alpha ,a_1,a_2\}$
model, became a coupled fermion after we added the fermionic sector.

\section{Conclusions}

\label{SecConcl}

We have explored the phenomenon of the exact solution of continuous
equations by their discretizations. We formulated the property of discrete
conformal invariance (DCI), and showed that any system of lattice equations
possessing DCI delivers exact solutions of its continuous limit. In this
sense, the conformal invariance is the cause of the exact solution property;
the continuous limit equations must also be conformally invariant (although
not all conformally invariant equations are exactly solved by any
discretizations). We found a class of lattice equations, based on Lie group
target space, with the property of DCI; their continuous limit corresponds
to a theory of WZW bosons. We also found a more general class of theories
based on semigroups describing scalar fields and fermions, which can be in
general nonlinearly coupled to each other.

In all these models, solutions to boundary value problems can be written
explicitly (see Eq.\ (\ref{ExactSol})) using the multiplication law of the
group or semigroup at hand. Expressed in this fashion through the boundary
conditions on a lightcone, the solutions are always algebraically factorized.

In case of the wave equation (\ref{WaveEqu}) with gauge constraints (\ref
{WEGaugeFix}), it was found that the discretized constraint equations (\ref
{DWEGaugeFix}) are also exactly solved by certain solutions of the
discretized wave equation, however they are not conformally invariant and,
correspondingly, the exact solutions of the equations (\ref{WaveEqu})$-$(\ref
{WEGaugeFix}) do not necessarily satisfy (\ref{DWEGaugeFix}) for any
discretization step $\Delta $. The ``compatibility'' of the discretized
equations (\ref{DWaveEqu}) and (\ref{DWEGaugeFix}) is perhaps due to the
simple algebraic form of the solutions (\ref{DWESol}).

The author is grateful to Alex Vilenkin for suggesting the problem and for
comments on the manuscript, and to Itzhak Bars, Arvind Borde, Oleg Gleizer,
Leonid Positselsky, and Washington Taylor for helpful and inspiring
discussions.

\section*{Appendix}

Here we show that if the binary operation $a*_cb$ on target space $V$ is not
only associative but is actually a group multiplication in some group $G$,
then the target space can be reparametrized so that the operation $*_c$
becomes, in terms of the group multiplication $*$, 
\begin{equation}
a*_cb=a*c^{-1}*b.
\end{equation}

By assumption, for each $c\in V$ there is a one-to-one map $g_c:V\rightarrow
G$ from the target space to the group $G$ such that 
\begin{equation}
g_c\left( a*_cb\right) =g_c\left( a\right) *g_c\left( b\right) .
\end{equation}
The first condition of (\ref{DCIConds}) can be written as 
\[
a*_b\left( b*_dc\right) =a*_dc, 
\]
or, if we take $g_b$ of both parts, 
\[
g_b\left( a\right) *g_b\left( b*_dc\right) =g_bg_d^{-1}\left( g_d\left(
a\right) *g_d\left( c\right) \right) . 
\]
If we now denote $g_d\left( a\right) \equiv x$ and $g_d\left( c\right)
\equiv y$, where $x$ and $y$ are elements of $G$, this relation becomes 
\begin{equation}
g_bg_d^{-1}\left( x\right) *g_bg_d^{-1}\left( g_d\left( b\right) *y\right)
=g_bg_d^{-1}\left( x*y\right) .  \label{rel2}
\end{equation}
This shows that $g_bg_d^{-1}$ is almost a group homomorphism, and we can
make it one if we define 
\[
h\left( b,d\right) \left( x\right) \equiv g_bg_d^{-1}\left( g_d\left(
b\right) *x\right) , 
\]
where $h\left( b,d\right) $ is a map $G\rightarrow G$. After this (\ref{rel2}%
) becomes 
\[
h\left( b,d\right) \left( x\right) *h\left( b,d\right) \left( y\right)
=h\left( b,d\right) \left( x*y\right) 
\]
for all $x,y\in G$. Now, obviously $h\left( a,a\right) $ is the identity
map, and $h\left( a,b\right) h\left( b,c\right) =h\left( a,c\right) $ for
all $a,b,c\in V$. This means that $h\left( a,b\right) $ can be expressed as 
\[
h\left( a,b\right) =\lambda \left( a\right) \left[ \lambda \left( b\right)
\right] ^{-1}, 
\]
where $\lambda \left( a\right) $ is an appropriately chosen homomorphism of $%
G$, and $\left[ \lambda \left( b\right) \right] ^{-1}$ is the inverse of the
homomorphism $\lambda \left( b\right) $. For example, we could choose an
arbitrary element $a_0\in V$ and define a map $\lambda :V\rightarrow $Hom$G$
by 
\[
\lambda \left( a\right) \equiv h\left( a,a_0\right) . 
\]

Now, if we modify the function $g_b\left( a\right) $ by a $\lambda $
transformation: 
\[
\tilde g_b\left( a\right) \equiv \left[ \lambda \left( b\right) \right]
^{-1}g_b\left( a\right) , 
\]
then we obtain 
\[
\tilde g_c\left( b\right) *\tilde g_b\left( a\right) =\tilde g_c\left(
a\right) , 
\]
which similarly means that $\tilde g_b\left( a\right) $ is of the form 
\[
\tilde g_b\left( a\right) =r\left( b\right) *\left[ r\left( a\right) \right]
^{-1}, 
\]
where $r:V\rightarrow G$ is an appropriately chosen 1-to-1 map, and $\left[
r\left( a\right) \right] ^{-1}$ is the group inverse of $r\left( a\right) $.

Finally, we can put the pieces together and find that 
\begin{equation}
F\left( a,b,c\right) =g_c^{-1}\left( g_c\left( a\right) *g_c\left( b\right)
\right) =r^{-1}\left( r\left( a\right) *\left[ r\left( c\right) \right]
^{-1}*r\left( b\right) \right) ,
\end{equation}
which is the desired result.


\begin{figure}[tbh]
\epsfysize 5 cm \epsffile{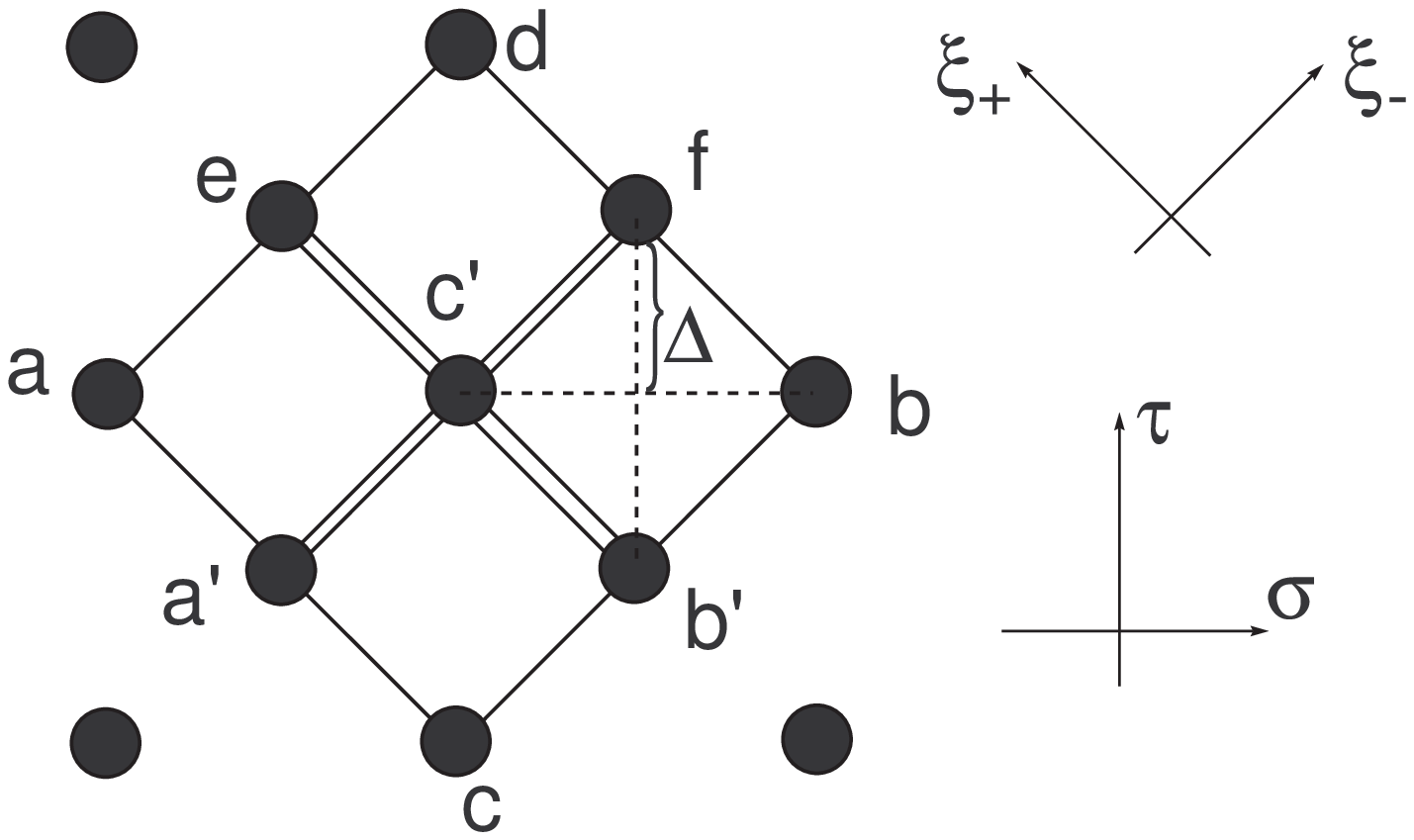} \label{NinePoints}
\caption{The discrete evolution equation applies to any four vertices of a
lattice rectangle.}
\end{figure}

\begin{figure}[tbh]
\epsfysize 5 cm \epsffile{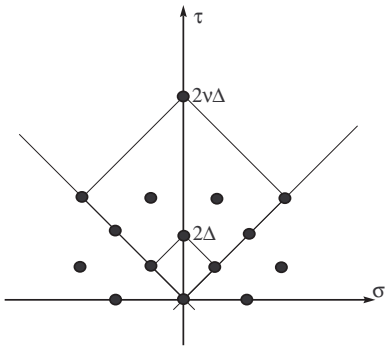} \label{PointN}
\caption{Comparison of the $1\times 1$ and $n\times n$ lattice squares.}
\end{figure}

\end{document}